\documentclass[conference]{IEEEtran}

\usepackage[utf8]{inputenc}
\usepackage[T1]{fontenc}

\usepackage{amsmath}
\usepackage{amssymb}
\usepackage{array}
\usepackage{dsfont}
\usepackage{manfnt}
\usepackage[caption=false]{subfig}
\usepackage{tabularx}
\usepackage{booktabs}
\usepackage{rotating}

\newcommand{\id}{\ensuremath{\text{id}}}
\newcommand{\cyc}{\ensuremath{\operatorname{cyc}}}
\newcommand{\type}{\ensuremath{\operatorname{type}}}
\newcommand{\size}{\ensuremath{\operatorname{size}}}
\newcommand{\trans}{\ensuremath{\operatorname{trans}}}

\newtheorem{example}{Example}
\newtheorem{lemma}{Lemma}
\newtheorem{theorem}{Theorem}

\newcommand{\bvand}{\ensuremath{\mathrel{\&}}}
\newcommand{\bvor}{\ensuremath{\mathrel{|}}}
\newcommand{\bvxor}{\ensuremath{\oplus}}

\newcommand{\mn}{\medskip\noindent}

\def\t(#1 #2){(#1,#2)}

\newcommand{\T}{\operatorname{T}}

\def\hang{\hangindent19pt}
\def\d@anger{\medbreak\begingroup\clubpenalty=10000
 \def\par{\endgraf\endgroup\medbreak} \noindent\hang\hangafter=-2
 \hbox to0pt{\hskip-\hangindent\dbend\hfill}\small}
\outer\def\danger{\d@anger}

\usepackage{tikz}
\usetikzlibrary{calc,matrix}

\tikzset{%
  >=stealth,
  font={\footnotesize}
}

\newcommand{\tikzSTG}{%
  \begin{tikzpicture}
    \draw[line width=0.30] (0.50,1.25) -- (1.00,1.25);
    \draw (0.40,1.25) node [left] {$x_1$};
    \draw (1.10,1.25) node [right] {$x_1$};
    \draw[line width=0.30] (0.50,0.75) -- (1.00,0.75);
    \draw (0.40,0.75) node [left] {$x_2$};
    \draw (1.10,0.75) node [right] {$x_2$};
    \draw[line width=0.30] (0.50,0.25) -- (1.00,0.25);
    \draw (0.40,0.25) node [left] {$x_3$};
    \draw (1.10,0.25) node [right] {$x_3\oplus (x_1 \lor x_2)$};
    \draw[line width=0.30] (0.75,0.25) -- (0.75,1.25);
    \draw[rounded corners=2pt,fill=white] ($(0.75,1.25)+(.15,.15)$) coordinate (a) rectangle ($(0.75,0.75)-(.15,.15)$) coordinate (b);
    \node at ($(a)!.5!(b)$) {$\lor$};
    \draw[line width=0.3] (0.75,0.25) circle (0.2) (0.75,0.05) -- (0.75,0.45);
  \end{tikzpicture}}

\newcommand{\tikzMPMCT}{%
  \begin{tikzpicture}
    \draw[line width=0.30] (0.50,1.25) -- (1.50,1.25);
    \draw (0.40,1.25) node [left] {$x_1$};
    \draw (1.60,1.25) node [right] {$x_1$};
    \draw[line width=0.30] (0.50,0.75) -- (1.50,0.75);
    \draw (0.40,0.75) node [left] {$x_2$};
    \draw (1.60,0.75) node [right] {$x_2$};
    \draw[line width=0.30] (0.50,0.25) -- (1.50,0.25);
    \draw (0.40,0.25) node [left] {$x_3$};
    \draw (1.60,0.25) node [right] {$x_3 \oplus (x_1\lor x_2)$};
    \draw[line width=0.30] (0.75,0.25) -- (0.75,1.25);
    \draw[fill=white] (0.75,1.25) circle (0.10);
    \draw[fill=white] (0.75,0.75) circle (0.10);
    \draw[line width=0.3] (0.75,0.25) circle (0.2) (0.75,0.05) -- (0.75,0.45);
    \draw[line width=0.30] (1.25,0.25) -- (1.25,0.25);
    \draw[line width=0.3] (1.25,0.25) circle (0.2) (1.25,0.05) -- (1.25,0.45);
  \end{tikzpicture}}

\newcommand{\tikzMCT}{%
  \begin{tikzpicture}
    \draw[line width=0.30] (0.50,1.25) -- (2.00,1.25);
    \draw (0.40,1.25) node [left] {$x_1$};
    \draw (2.10,1.25) node [right] {$x_1$};
    \draw[line width=0.30] (0.50,0.75) -- (2.00,0.75);
    \draw (0.40,0.75) node [left] {$x_2$};
    \draw (2.10,0.75) node [right] {$x_2$};
    \draw[line width=0.30] (0.50,0.25) -- (2.00,0.25);
    \draw (0.40,0.25) node [left] {$x_3$};
    \draw (2.10,0.25) node [right] {$x_3\oplus (x_1 \lor x_2)$};
    \draw[line width=0.30] (0.75,0.25) -- (0.75,1.25);
    \draw[fill] (0.75,1.25) circle (0.10);
    \draw[fill] (0.75,0.75) circle (0.10);
    \draw[line width=0.3] (0.75,0.25) circle (0.2) (0.75,0.05) -- (0.75,0.45);
    \draw[line width=0.30] (1.25,0.25) -- (1.25,1.25);
    \draw[fill] (1.25,1.25) circle (0.10);
    \draw[line width=0.3] (1.25,0.25) circle (0.2) (1.25,0.05) -- (1.25,0.45);
    \draw[line width=0.30] (1.75,0.25) -- (1.75,0.75);
    \draw[fill] (1.75,0.75) circle (0.10);
    \draw[line width=0.3] (1.75,0.25) circle (0.2) (1.75,0.05) -- (1.75,0.45);
  \end{tikzpicture}}

\newcommand{\tikzAncConstr}{%
\begin{tikzpicture}
\draw[line width=0.30] (0.50,2.75) -- (5.00,2.75);
\draw (0.40,2.75) node [left] {$x_1$};
\draw (5.10,2.75) node [right] {$y_1$};
\draw[line width=0.30] (0.50,2.25) -- (5.00,2.25);
\draw (0.40,2.25) node [left] {$x_2$};
\draw (5.10,2.25) node [right] {$y_2$};
\draw[line width=0.30] (0.50,1.25) -- (5.00,1.25);
\draw (0.40,1.25) node [left] {$x_{n-1}$};
\draw (5.10,1.25) node [right] {$y_{n-1}$};
\draw[line width=0.30] (0.50,0.75) -- (5.00,0.75);
\draw (0.40,0.75) node [left] {$x_n$};
\draw (5.10,0.75) node [right] {$y_n$};
\draw[line width=0.30] (0.50,0.25) -- (5.00,0.25);
\draw (0.40,0.25) node [left] {$0$};
\draw (5.10,0.25) node [right] {$0$};
\draw[line width=0.30] (0.75,0.25) -- (0.75,2.75);
\draw[fill] (0.75,2.75) circle (0.10); \draw[fill=white] ($(0.75,2.75)+(45:.1)$) arc (45:225:0.10);
\draw[fill] (0.75,2.25) circle (0.10); \draw[fill=white] ($(0.75,2.25)+(45:.1)$) arc (45:225:0.10);
\draw[fill] (0.75,1.25) circle (0.10); \draw[fill=white] ($(0.75,1.25)+(45:.1)$) arc (45:225:0.10);
\draw[line width=0.3] (0.75,0.25) circle (0.2) (0.75,0.05) -- (0.75,0.45);
\draw[line width=0.30] (1.25,0.25) -- (1.25,2.75);
\draw[fill] (1.25,2.75) circle (0.10); \draw[fill=white] ($(1.25,2.75)+(45:.1)$) arc (45:225:0.10);
\draw[fill] (1.25,2.25) circle (0.10); \draw[fill=white] ($(1.25,2.25)+(45:.1)$) arc (45:225:0.10);
\draw[fill] (1.25,1.25) circle (0.10); \draw[fill=white] ($(1.25,1.25)+(45:.1)$) arc (45:225:0.10);
\draw[line width=0.3] (1.25,0.25) circle (0.2) (1.25,0.05) -- (1.25,0.45);
\draw[line width=0.30] (1.75,0.25) -- (1.75,2.75);
\draw[fill] (1.75,2.75) circle (0.10); \draw[fill=white] ($(1.75,2.75)+(45:.1)$) arc (45:225:0.10);
\draw[fill] (1.75,2.25) circle (0.10); \draw[fill=white] ($(1.75,2.25)+(45:.1)$) arc (45:225:0.10);
\draw[fill] (1.75,1.25) circle (0.10); \draw[fill=white] ($(1.75,1.25)+(45:.1)$) arc (45:225:0.10);
\draw[line width=0.3] (1.75,0.25) circle (0.2) (1.75,0.05) -- (1.75,0.45);
\draw[line width=0.30] (2.25,0.25) -- (2.25,2.75);
\draw[fill] (2.25,2.75) circle (0.10); \draw[fill=white] ($(2.25,2.75)+(45:.1)$) arc (45:225:0.10);
\draw[fill] (2.25,2.25) circle (0.10); \draw[fill=white] ($(2.25,2.25)+(45:.1)$) arc (45:225:0.10);
\draw[fill] (2.25,1.25) circle (0.10); \draw[fill=white] ($(2.25,1.25)+(45:.1)$) arc (45:225:0.10);
\draw[line width=0.3] (2.25,0.25) circle (0.2) (2.25,0.05) -- (2.25,0.45);
\draw[line width=0.30] (2.75,0.25) -- (2.75,0.75);
\draw[fill] (2.75,0.25) circle (0.10);
\draw[line width=0.3] (2.75,0.75) circle (0.2) (2.75,0.55) -- (2.75,0.95);
\draw[line width=0.30] (3.25,0.25) -- (3.25,2.75);
\draw[fill] (3.25,2.75) circle (0.10); \draw[fill=white] ($(3.25,2.75)+(45:.1)$) arc (45:225:0.10);
\draw[fill] (3.25,2.25) circle (0.10); \draw[fill=white] ($(3.25,2.25)+(45:.1)$) arc (45:225:0.10);
\draw[fill] (3.25,1.25) circle (0.10); \draw[fill=white] ($(3.25,1.25)+(45:.1)$) arc (45:225:0.10);
\draw[line width=0.3] (3.25,0.25) circle (0.2) (3.25,0.05) -- (3.25,0.45);
\draw[line width=0.30] (3.75,0.25) -- (3.75,2.75);
\draw[fill] (3.75,2.75) circle (0.10); \draw[fill=white] ($(3.75,2.75)+(45:.1)$) arc (45:225:0.10);
\draw[fill] (3.75,2.25) circle (0.10); \draw[fill=white] ($(3.75,2.25)+(45:.1)$) arc (45:225:0.10);
\draw[fill] (3.75,1.25) circle (0.10); \draw[fill=white] ($(3.75,1.25)+(45:.1)$) arc (45:225:0.10);
\draw[line width=0.3] (3.75,0.25) circle (0.2) (3.75,0.05) -- (3.75,0.45);
\draw[line width=0.30] (4.25,0.25) -- (4.25,2.75);
\draw[fill] (4.25,2.75) circle (0.10); \draw[fill=white] ($(4.25,2.75)+(45:.1)$) arc (45:225:0.10);
\draw[fill] (4.25,2.25) circle (0.10); \draw[fill=white] ($(4.25,2.25)+(45:.1)$) arc (45:225:0.10);
\draw[fill] (4.25,1.25) circle (0.10); \draw[fill=white] ($(4.25,1.25)+(45:.1)$) arc (45:225:0.10);
\draw[line width=0.3] (4.25,0.25) circle (0.2) (4.25,0.05) -- (4.25,0.45);
\draw[line width=0.30] (4.75,0.25) -- (4.75,2.75);
\draw[fill] (4.75,2.75) circle (0.10); \draw[fill=white] ($(4.75,2.75)+(45:.1)$) arc (45:225:0.10);
\draw[fill] (4.75,2.25) circle (0.10); \draw[fill=white] ($(4.75,2.25)+(45:.1)$) arc (45:225:0.10);
\draw[fill] (4.75,1.25) circle (0.10); \draw[fill=white] ($(4.75,1.25)+(45:.1)$) arc (45:225:0.10);
\draw[line width=0.3] (4.75,0.25) circle (0.2) (4.75,0.05) -- (4.75,0.45);
\draw (0.6,3) -- ++(right:1.3) -- ++(down:2pt) (0.6,3) -- ++(down:2pt);
\node[above] at (1.25,3) {$r$};
\node[above] at (2.25,2.9) {$g$};
\draw (3.6,3) -- ++(right:1.3) -- ++(down:2pt) (3.6,3) -- ++(down:2pt);
\node[above] at (4.25,3) {$r$};
\node[above] at (3.25,2.9) {$g$};
\draw (0.5,-0.2) -- ++(right:4.5) -- ++(up:2pt) (0.5,-0.2) -- ++(up:2pt);
\node[below] at (2.75,-0.2) {$h$};
\end{tikzpicture}}

\newcommand{\tikzVConstr}{%
\begin{tikzpicture}
\draw[line width=0.30] (0.50,2.25) -- (4.00,2.25);
\draw (0.40,2.25) node [left] {$x_1$};
\draw (4.10,2.25) node [right] {$y_1$};
\draw[line width=0.30] (0.50,1.75) -- (4.00,1.75);
\draw (0.40,1.75) node [left] {$x_2$};
\draw (4.10,1.75) node [right] {$y_2$};
\draw[line width=0.30] (0.50,0.75) -- (4.00,0.75);
\draw (0.40,0.75) node [left] {$x_{n-1}$};
\draw (4.10,0.75) node [right] {$y_{n-1}$};
\draw[line width=0.30] (0.50,0.25) -- (4.00,0.25);
\draw (0.40,0.25) node [left] {$x_n$};
\draw (4.10,0.25) node [right] {$y_n$};
\draw[line width=0.30] (0.75,0.25) -- (0.75,2.25);
\draw[fill] (0.75,2.25) circle (0.10); \draw[fill=white] ($(0.75,2.25)+(45:.1)$) arc (45:225:0.10);
\draw[fill] (0.75,1.75) circle (0.10); \draw[fill=white] ($(0.75,1.75)+(45:.1)$) arc (45:225:0.10);
\draw[fill] (0.75,0.75) circle (0.10); \draw[fill=white] ($(0.75,0.75)+(45:.1)$) arc (45:225:0.10);
\draw[fill=white] ($(0.75,0.25)+(135:.3)$) rectangle ($(0.75,0.25)+(315:.3)$); \node at (0.75,0.25) {$V$};
\draw[line width=0.30] (1.25,0.25) -- (1.25,2.25);
\draw[fill] (1.25,2.25) circle (0.10); \draw[fill=white] ($(1.25,2.25)+(45:.1)$) arc (45:225:0.10);
\draw[fill] (1.25,1.75) circle (0.10); \draw[fill=white] ($(1.25,1.75)+(45:.1)$) arc (45:225:0.10);
\draw[fill] (1.25,0.75) circle (0.10); \draw[fill=white] ($(1.25,0.75)+(45:.1)$) arc (45:225:0.10);
\draw[fill=white] ($(1.25,0.25)+(135:.3)$) rectangle ($(1.25,0.25)+(315:.3)$); \node at (1.25,0.25) {$V$};
\draw[line width=0.30] (1.75,0.25) -- (1.75,2.25);
\draw[fill] (1.75,2.25) circle (0.10); \draw[fill=white] ($(1.75,2.25)+(45:.1)$) arc (45:225:0.10);
\draw[fill] (1.75,1.75) circle (0.10); \draw[fill=white] ($(1.75,1.75)+(45:.1)$) arc (45:225:0.10);
\draw[fill] (1.75,0.75) circle (0.10); \draw[fill=white] ($(1.75,0.75)+(45:.1)$) arc (45:225:0.10);
\draw[fill=white] ($(1.75,0.25)+(135:.3)$) rectangle ($(1.75,0.25)+(315:.3)$); \node at (1.75,0.25) {$V$};
\draw[line width=0.30] (2.25,0.25) -- (2.25,2.25);
\draw[fill] (2.25,2.25) circle (0.10); \draw[fill=white] ($(2.25,2.25)+(45:.1)$) arc (45:225:0.10);
\draw[fill] (2.25,1.75) circle (0.10); \draw[fill=white] ($(2.25,1.75)+(45:.1)$) arc (45:225:0.10);
\draw[fill] (2.25,0.75) circle (0.10); \draw[fill=white] ($(2.25,0.75)+(45:.1)$) arc (45:225:0.10);
\draw[line width=0.3] (2.25,0.25) circle (0.2) (2.25,0.05) -- (2.25,0.45);
\draw[line width=0.30] (2.75,0.25) -- (2.75,2.25);
\draw[fill] (2.75,2.25) circle (0.10); \draw[fill=white] ($(2.75,2.25)+(45:.1)$) arc (45:225:0.10);
\draw[fill] (2.75,1.75) circle (0.10); \draw[fill=white] ($(2.75,1.75)+(45:.1)$) arc (45:225:0.10);
\draw[fill] (2.75,0.75) circle (0.10); \draw[fill=white] ($(2.75,0.75)+(45:.1)$) arc (45:225:0.10);
\draw[fill=white] ($(2.75,0.25)+(135:.3)$) rectangle ($(2.75,0.25)+(315:.3)$); \node at (2.75,0.25) {$V$};
\draw[line width=0.30] (3.25,0.25) -- (3.25,2.25);
\draw[fill] (3.25,2.25) circle (0.10); \draw[fill=white] ($(3.25,2.25)+(45:.1)$) arc (45:225:0.10);
\draw[fill] (3.25,1.75) circle (0.10); \draw[fill=white] ($(3.25,1.75)+(45:.1)$) arc (45:225:0.10);
\draw[fill] (3.25,0.75) circle (0.10); \draw[fill=white] ($(3.25,0.75)+(45:.1)$) arc (45:225:0.10);
\draw[fill=white] ($(3.25,0.25)+(135:.3)$) rectangle ($(3.25,0.25)+(315:.3)$); \node at (3.25,0.25) {$V$};
\draw[line width=0.30] (3.75,0.25) -- (3.75,2.25);
\draw[fill] (3.75,2.25) circle (0.10); \draw[fill=white] ($(3.75,2.25)+(45:.1)$) arc (45:225:0.10);
\draw[fill] (3.75,1.75) circle (0.10); \draw[fill=white] ($(3.75,1.75)+(45:.1)$) arc (45:225:0.10);
\draw[fill] (3.75,0.75) circle (0.10); \draw[fill=white] ($(3.75,0.75)+(45:.1)$) arc (45:225:0.10);
\draw[fill=white] ($(3.75,0.25)+(135:.3)$) rectangle ($(3.75,0.25)+(315:.3)$); \node at (3.75,0.25) {$V$};

\node[above] at (2.25,2.3) {$g$};
\draw (0.6,2.5) -- ++(right:1.3) -- ++(down:2pt) (0.6,2.5) -- ++(down:2pt);
\draw (2.6,2.5) -- ++(right:1.3) -- ++(down:2pt) (2.6,2.5) -- ++(down:2pt);
\draw (1.25,2.5) to[out=80,in=100] node[above] {$r$} (3.25,2.5);
\draw (0.5,-0.2) -- ++(right:3.5) -- ++(up:2pt) (0.5,-0.2) -- ++(up:2pt);
\node[below] at (2.25,-0.2) {$h$};
\end{tikzpicture}}

\title{Self-Inverse Functions and Palindromic Circuits}
\author{%
  \IEEEauthorblockN{Mathias Soeken$^{1,2}$ \qquad Michael Kirkedal Thomsen$^1$
    \qquad Gerhard W.\ Dueck$^3$ \qquad D.\ Michael Miller$^4$}
  \IEEEauthorblockA{%
    $^1$ Department of Mathematics and Computer Science, University of Bremen,
    Germany \\
    $^2$ Cyber-Physical Systems, DFKI GmbH, Bremen, Germany \\
    $^3$ Faculty of Computer Science, University of New Brunswick, Fredericton,
    NB, Canada \\
    $^4$ Department of Computer Science, University of Victoria, Victoria, BC,
    Canada \\
    \{msoeken,kirkedal\}@cs.uni-bremen.de \qquad gdueck@unb.ca \qquad mmiller@uvic.ca
  }
}

\begin{document}

\maketitle

\begin{abstract}
  We investigate the subclass of reversible functions that are self-inverse and
  relate them to reversible circuits that are equal to their reverse circuit,
  which are called \emph{palindromic circuits}.  We precisely determine which
  self-inverse functions can be realized as a palindromic circuit.  For those
  functions that cannot be realized as a palidromic circuit, we find alternative
  palindromic representations that require an extra circuit line or quantum
  gates in their construction. Our analyses make use of involutions in the
  symmetric group $S_{2^n}$ which are isomorphic to self-inverse reversible
  function on $n$ variables.
\end{abstract}

\section{Introduction}

While the reversible circuit model has seen many practical applications (e.g.,
logic designs~\cite{KPF:08,CDKM:05,DeVos:10}, reversible logic
synthesis~\cite{MDM:07,GWDD:09,SWH+:12}), the theoretical aspects of the logic
circuit model have received much less attention. This is, in it self, not a
hindrance to the usage of the logic model in the aforementioned applications,
but it does limit our understanding and therefore the possibility to implement
the applications most efficiently.

In this paper we investigate the relationship between (reversible) self-inverse
functions (involutions) and reversible \emph{palindromic} circuits. By a
palindromic circuit we mean a reversible circuit generated from gates and serial
circuit composition (no parallel composition) that is identical when reading it
from left and right.

Looking at reversible circuit as permutations is not a novel idea. This duality
has been used for reversible logic synthesis~\cite{SPMH:03,VR:05} but also as
theoretical foundation for reversible logic analysis~\cite{SVJ:99,ASTD:14}.
Though the many results have shown these to be interesting approaches, we will
take a different approach for this work. To get a deep understanding of
palindromic circuits, we define which permutations (defined as transpositions in
the cycle notation) are equivalent to \emph{mixed-polarity multiply-controlled
  Toffoli gates} (MPMCT). For this purpose we exploit general theorems about
permutations.

The authors in \cite{KS:11} have coined the term palindromic circuits and also
related them to self-inverse functions.  They have shown that there are some
self-inverse functions that can be realized as a palindromic circuit and argued
that for some no such realization can be found.  In this paper we precisely
determine which self-inverse functions can be realized as a palindromic
circuit. For those functions that cannot be realized as a palindromic circuit,
we find alternative palindromic representations that require an extra circuit
line or quantum gates in their construction.  In~\cite{AS:03} palindromic
circuits have been used in an optimization technique for quantum circuits.

The paper is organized as follows.  Basic notations and definitions for
permutations and reversible circuits are described in the next section.
Section~\ref{sec:self-inverse-revers} discusses properties of self-inverse
reversible functions and shows how MPMCT gates can be derived from
transpositions.  Section~\ref{sec:palindromic-circuits} introduces palindromic
circuits and determines the subclass of self-inverse functions that can be
realized as a palindromic circuit.  Section~\ref{sec:altern-constr} illustrates
alternative constructions for palindromic circuits that can realize all
self-inverse function and Section~\ref{sec:conclusions} concludes the paper.

\section{Preliminaries}
\subsection{Basic Notation and Definitions}
Applying the bit-wise operations `\bvand', `\bvor', and `\bvxor' to non-negative
numbers is interpreted as applying them to their unsigned bit-wise expansion.
The operation `$\nu$' is the sideways sum and counts the number of ones in a
bit-string or in the bit-wise expansion of a non-negative number.  The
\emph{double factorial} $n!! = \prod_{i=0}^{\lceil n/2\rceil -1}(n-2i)$ is the
product of all integers from $1$ to $n$ that have the same polarity as $n$.  For
a non-negative number $n$, an \emph{integer partition} $n$ is a sequence
$\mu = (\mu_1, \mu_2, \dots, \mu_k)$ such that
$\mu_1 \ge \mu_2 \ge \cdots \ge \mu_k$ and $\mu_1 + \mu_2 + \cdots + \mu_k = n$.

\subsection{Permutations}
Permutations are elements from the symmetric group $S_n$ i.e.~bijections over
the set $\{ 0, 1, \dots, n-1\}$.  We chose to have $0$ as the lowest permutation
index, in contrast to the conventional definition, as this makes computation
with respect to reversible functions and gates easier.  Several notations are
used for permutations.  Given a permutation $\pi \in S_n$ its \emph{two-line
  form} representation is
\begin{equation}
  \label{eq:two-line}
  \begin{pmatrix}
    i_1      & i_2      & \cdots & i_n      \\
    \pi(i_1) & \pi(i_2) & \cdots & \pi(i_n)
  \end{pmatrix}
\end{equation}
in which all indexes are written in the first line and its function values with
respect to $\pi$ in the second line.  The order of indexes in the first line is
arbitrary, however, if we have $i_1 < i_2 < \cdots < i_n$ we can omit the first
line and have the \emph{one-line form} representation
\begin{equation}
  \label{eq:one-line}
  \begin{pmatrix}
    \pi(i_1) & \pi(i_2) & \cdots & \pi(i_n)
  \end{pmatrix}.
\end{equation}
A permutation can be partitioned into \emph{cycles} $(i_1 , i_2 , \cdots , i_k)$
such that $\pi(i_j) = i_{j+1}$ for $j < k$ and $\pi(i_k) = i_1$.  The order of
cycles and the starting value inside a cycle do not change the permutation.  A
cycle of length $1$ is called a \emph{fixpoint} and a cycle of length $2$ is
called a \emph{transposition}.  Fixpoints are usually omitted in the cyclic
representation.  Given a permutation $\pi \in S_n$ in cyclic notation, we refer
to the number of cycles (including fixpoints) as $\cyc(\pi)$.  Also let
$\type(\pi)$ be the list of sizes of these cycles, including repetitions,
written in decreasing order, i.e., $\type(\pi)$ is an integer partition of $n$.
The permutation that represents the identity is denoted $\pi_\id$.

\begin{example}
  Let $\pi \in S_8$ be a permutation with two-line form
  $\left(\begin{smallmatrix}
    0 & 7 & 2 & 4 & 6 & 5 & 3 & 1 \\
    4 & 7 & 6 & 3 & 5 & 1 & 0 & 2
   \end{smallmatrix}\right)$.  The two-line form in which the first line is
 ordered is
  $\left(\begin{smallmatrix}
    0 & 1 & 2 & 3 & 4 & 5 & 6 & 7 \\
    4 & 2 & 6 & 0 & 3 & 1 & 5 & 7
  \end{smallmatrix}\right)$ from which we can immediately extract the one-line
form $\begin{pmatrix} 4 & 2 & 6 & 0 & 3 & 1 & 5 & 7 \end{pmatrix}$.  The cyclic
representation of $\pi$ is $(0 , 4 , 3)(1 , 2 , 6 , 5)(7)$.  We
have $\cyc(\pi) = 3$ and $\type(\pi) = (4,3,1)$.  There are no transpositions in
the cyclic representation and the only fixpoint is $7$.
\end{example}

\mn The notion of type can be used to partition permutations into conjugacy
classes.  For this purpose, we review two well-known lemmas.

\begin{lemma}
  \label{lem:same-type}
  For all permutations $\pi,\sigma \in S_n$ we have $\type(\sigma \circ \pi
  \circ \sigma^{-1}) = \type(\pi)$.
\end{lemma}

\begin{IEEEproof}
  We show that if
  \[\pi = (i_1 , i_2 , \dots)(j_1 , j_2 , \dots) \cdots\]
then
\[ \sigma\pi\sigma^{-1} = (\sigma(i_1), \sigma(i_2), \dots)(\sigma(j_1),
\sigma(j_2), \dots) \cdots. \]
We first assume that $\cyc(\pi) = 1$, i.e., $\pi = (i_1, i_2, \dots, i_k)$ and
show that $\sigma\pi\sigma^{-1}$ and
$\pi' = (\sigma(i_1), \sigma(i_2), \dots, \sigma(i_k))$ are equal by proving
that both have the same effect on $x \in \{1, 2, \dots, n\}$.  First assume that
$x = \sigma(i_s)$ for some $1 \le s \le k$.  Then
\[\sigma\pi\sigma^{-1}(x) = \sigma\pi\sigma^{-1}\sigma(i_s) = \sigma\pi(i_s) =
\sigma(i_{(s+1)\mathop{\%}k}) = \sigma(x).\]
If $x \neq \sigma(i_s)$ for any $s$, then $\sigma(x) = \sigma^{-1}(x) = x$ and
$\pi$ fixes $\sigma^{-1}(x)$.  The general form for multiple cycles follows from
conjugation being a homomorphism.  See also~\cite{Loehr:11}.
\end{IEEEproof}

\begin{lemma}
  \label{lem:conj-class}
  Let $\pi, \pi' \in S_n$ such that $\type(\pi) = \type(\pi')$.  Then there
  exists a permutation $\sigma$ such that $\pi = \sigma \circ \pi' \circ
  \sigma^{-1}$.
\end{lemma}

\begin{IEEEproof}
  When writing $\pi$ atop $\pi'$ such that the size of cycles match one obtains
  $\sigma$ in two-line form.  Due to ordering of same sized cycles and elements
  in cycles several permutations for $\sigma$ can be obtained, unless
  $\pi=\pi_{\rm id}$.
\end{IEEEproof}

The \emph{inverse} $\pi^{-1}$ of a permutation $\pi$ is found by swapping the
first and second line in its two-line form.  A permutation $\pi$ is called an
\emph{involution} if $\pi = \pi^{-1}$.  (Sometimes, $\pi$ is also called
self-inverse or self-conjugate.)

\begin{lemma}
  Let $\pi$ be an involution. Then, the cycle representation of $\pi$ consists
  only of transpositions and fixpoints.
\end{lemma}
\begin{IEEEproof}
  The cycle representation is unique when disregarding order of cycles and order
  of elements within cycles.  Assume that the cycle representation of $\pi$
  consists of a cycle $(i_1, i_2, \ldots , i_k)$ with $k>2$.  Then $\pi^{-1}$
  consists of the cycle $(i_k , \ldots , i_2 , i_1)$ and hence $\pi\neq
  \pi^{-1}$.
\end{IEEEproof}

Given an involution $\pi \in S_n$, let $\size(\pi)$ be the number of
transpositions in $\pi$.  Further, let $\trans(\pi)$ be the set of
transpositions in $\pi$.  We have $|\trans(\pi)|=\size(\pi)$ and
$\cyc(\pi) = n - \size(\pi)$.  Given a set of permutations $\Pi$, we define
\begin{equation}
  \label{eq:power-set-permutation}
  \mathcal{P}_\circ(\Pi) = \{\pi_1 \circ \pi_2 \circ \cdots \circ \pi_k \mid
    \{\pi_1, \pi_2, \ldots, \pi_k\} \subseteq \Pi \},
\end{equation}
referred to as the \emph{power set of permutations}.

\subsection{Reversible Circuits}
Reversible functions can be realized by reversible circuits that consist of at
least~$n$ lines and are constructed as cascades of reversible gates that belong
to a certain universal gate library. The most common gate library consists of Toffoli
gates or single-target gates.

Given a set of variables $X=\{x_1, \dots, x_n\}$, a \emph{reversible
  single-target gate} $\T_g(t)$ realizes a reversible functions on $n$ lines
that inverts the variable on the \emph{target line} $t \in X$ if and only if the
\emph{control function} $g$ evaluates to true, where $g$ is a Boolean function
with input variables $X \setminus \{t\}$.  Only line $t$ is updated.  The domain
of $g$ can be smaller than $X \setminus \{t\}$.

\begin{figure}[t]
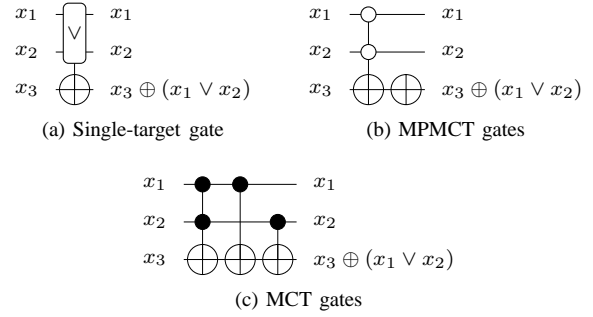

  \centering
  \subfloat[Single-target gate]{\tikzSTG} \hfil
  \subfloat[MPMCT gates]{\tikzMPMCT}      \hfil
  \subfloat[MCT gates]{\tikzMCT}
  \caption{Reversible circuits that update $x_3$ with $x_1\lor x_2$}
  \label{fig:gates}
\end{figure}

\begin{example}
  Fig.~\ref{fig:gates}(a) shows the graphical notation of a single-target gate
  $\T_{x_1\lor x_2}(x_3)$ with control function $x_1 \lor x_2$ and target line
  $x_3$.
\end{example}

\mn There exist $n \cdot 2^{2^{n-1}}$ different single-target gates on $n$
lines, since for each target line one can choose from $2^{2^{n-1}}$ Boolean
functions over $n-1$ variables.  If the control function is $\bot$
(\emph{false}), the target line is never inverted and is therefore omitted from
the circuit representation.

\emph{Mixed-polarity multiple-control Toffoli~(MPMCT) gates} are a subset of the
single-target gates in which the control function~$g$ is $\top$ (\emph{true}) or
can be represented as one product term consisting of positive and negative
literals over $X \setminus \{t\}$.  As notation we use $\T(C, t)$ where $C$ is
the set of literals in the product term.  If $g=\top$, $C$ is empty and the gate
is a \emph{Not} gate on line $t$.  The affected lines in $C$ are referred to as
\emph{control lines} and a line $x_i$ is called \emph{positive} if $x_i \in C$
and \emph{negative} if $\bar x_i \in C$.  \emph{Multiple-control Toffoli
  gates~(MCT)} are a subset of MPMCT gates in which the product terms can only
consist of positive literals.

\begin{example}
  Figs.~\ref{fig:gates}(b) and (c) show circuits consisting of MPMCT and MCT
  gates, respectively.  The gates in Fig.~\ref{fig:gates}(b) are
  $\T(\{\bar x_1, \bar x_2\}, x_3)$ and $\T(\emptyset, x_3)$.  The gates in
  Fig.~\ref{fig:gates}(c) are $\T(\{x_1, x_2\}, x_3)$, $\T(\{x_1\}, x_3)$, and
  $\T(\{x_2\}, x_3)$.
\end{example}

\section{Self-inverse Reversible Functions}
\label{sec:self-inverse-revers}
A reversible function $f$ on $n$ variables is called \emph{self-inverse} if
$f(f(x))=x$ for all input assignments $x$, or in other words if $f=f^{-1}$.  To
better understand these functions, it helps a lot to investigate the respective
permutations that are represented by the reversible functions, i.e., elements
from the symmetric group $S_{2^n}$.  Then, self-inverse functions correspond to
involutions.  The permutation matrix of an involution is symmetric.

\subsection{Reversible Gates}
The reversible gates that have been introduced in the previous section are
obviously self-inverse.  We are interested in transpositions that occur in
permutation representations of reversible gates that act on $n$ circuit lines.
Involutions whose number of transpositions is a power of 2 are playing a central
role when describing such gates.  For this purpose, we define
\begin{equation}
  \label{eq:involutions-k}
  I_n^k = \{ \pi \in S_{2^n} \mid \text{$\pi = \pi^{-1}$ and $\size(\pi)=2^{k-1}$} \}
\end{equation}
to be the set of all involutions over $2^n$ elements of size $2^{k-1}$ for
$1\le k\le n$.  We also define
\begin{equation}
  \label{eq:involutions}
  I_n = \bigcup_{k=1}^n I_n^k
\end{equation}
to be the set of all involutions which size is a power of 2.

Since the introduced reversible gates only change at most one bit at a time,
the occurring transpositions must be of the form $\t(a b)$ such that the
hamming distance of the binary expansions of $a = a_n \dots a_2a_1$ and $b = b_n
\dots b_2b_1$ is 1.  Let us refer to all of this transpositions as the set
$H_n$, i.e.,
\begin{equation}
  \label{eq:hn}
  H_n = \{ \t(a b) \mid \nu(a \oplus b) = 1 \}.
\end{equation}
First note that each transposition $\t(a b) \in H_n$ corresponds to one fully
controlled MPMCT gate.  It acts on line $i$ where $i$ is the single index for
which $a_i \neq b_i$.  The polarity of the controls is chosen according to the
other bits.  We have $|H_n|=\frac{2^n \cdot n}{2}$, because one has $2^n$
choices for $a$ and then $n$ choices for $b$ remain.  Since transposition is
commutative, the product needs to be halved.  Note that this number corresponds
to the number of fully controlled MPMCT gates $n \cdot 2^{n-1}$, i.e., one has
$n$ choices for the target and then each remaining line can be either positively
or negatively controlled.

Based on this observation we partition the set $H_n$ into $n$ sets
$H_{n,1}, H_{n,2}, \dots, H_{n,n}$ such that
\begin{equation}
  \label{eq:hni}
  H_{n,i} = \{ \t(a b) \in H_n \mid a \oplus b = 2^{i-1} \}
\end{equation}
contains all transpositions in which the components differ in their $i$-th bit.
Let $g$ be a single-target gate that acts on the $i$-th line and $\pi_g$ its
permutation representation, then $\trans(\pi_g)\subseteq H_{n,i}$.  But also the
reverse holds, i.e.~by selecting a subset of $H_{n,i}$ one finds a set of
transpositions that corresponds to a single target gate that acts on the $i$-th
line.  This can be easily found by counting as $|H_{n,i}| = 2^{n-1}$ and thus
there exist $2^{2^{n-1}}$ subsets which equals the number of Boolean functions
on $n-1$ variables.

\begin{example}
  For $n=3$, the following $12$ transpositions can be used to form gates that
  act on three circuit lines (brackets and commas for the sets have been removed
  for clarity):
  \begin{align*}
    H_{3,1} & = \t(0 1) \t(2 3) \t(4 5) \t(6 7) \\
    H_{3,2} & = \t(0 2) \t(1 3) \t(4 6) \t(5 7) \\
    H_{3,3} & = \t(0 4) \t(1 5) \t(2 6) \t(3 7) \\
  \end{align*}
\end{example}

\mn From all the subsets in $H_{n,i}$, there are $3^{n-1}$ subsets that
represent an MPMCT gate, since $3^{n-1}$ is the number of product terms over
$n-1$ variables.  The question is how these subsets are characterized.  One can
easily see that a MPMCT gate is represented by $2^{k-1}$ transpositions, where
$n-k$ is the number of control lines, i.e., there are $k-1$ empty lines.  But by
simply counting we see that not all subsets which size is a power 2 can
represent an MPMCT gate.  We need to select $2^{k-1}$ transpositions such that
the number of positions in which the overall bits of the binary expansions
differ is $k$, in other words, $\pi \in I_n^k$ represents an MPMCT gate, if and
only if $\nu p = k$ with
\begin{equation}
  \label{eq:gate-in-ink}
  p = \bigoplus \{a \oplus b \mid \t(a b) \in \trans(\pi)\}.
\end{equation}

\begin{example}
  As an example, an MPMCT gate with one control line in a circuit of 3 lines,
  i.e.~$k=2$, can be characterized by two transpositions from $H_{3,i}$ for some
  $i$.  The two transpositions $\t(4 5)\t(6 7)$ are a valid choice since their
  binary expansions $100$, $101$, $110$, and $111$ differ in 2 positions (last
  two bits).  The two transpositions $\t(2 3)\t(4 5)$, however, do not form an
  MPMCT gate since their binary expansions $010$, $011$, $100$, and $101$ differ
  in 3 positions.
\end{example}

\mn With all these observations, we finally define the set $G_n \subseteq I_n$
as the set of all permutations that represent MPMCT gates over $n$ lines
according to \eqref{eq:gate-in-ink}, based on which
\begin{equation}
  \label{eq:gate-by-line}
  G_{n,i} = G_n \cap \mathcal{P}_\circ(H_{n,i})
\end{equation}
is the set of MPMCT gates acting on line $i$ and
\begin{equation}
  \label{eq:gate-by-controls}
  G_n^k = G_n \cap I_n^k
\end{equation}
is the set of all MPMCT gates with $n-k$ control lines.  From these sets one
can derive
\begin{equation}
  \label{eq:gate-by-lines-and-controls}
  G_{n,i}^k = G_{n,i} \cap G_n^k
\end{equation}
as the set of all MPMCT gates with $n-k$ controls acting on line~$i$.

\subsection{Counting Self-Inverse Functions}
\begin{table*}[t]
  \centering
  \caption{Counting reversible functions}
  \label{tab:counting}
  \begin{tabular}{lrrrrr}
    \toprule
    & $n = 1$ & $n = 2$ & $n = 3$ & $n = 4$ & $n = 5$ \\
    \midrule
reversible & 2 & 24 & 40,240 & 20,922,789,888,000 & 263,130,836,933,693,530,167,218,012,160,000,000 \\
self-inverse & 2 & 10 & 764 & 46,206,736 & 22,481,059,424,730,750,976 \\
self-inverse (palindromic, $|I_n|$) & 1 & 9 & 343 & 3,383,955 & 193,117,190,044,580,256 \\
single-target gate & 2 & 7 & 46 & 1,021 & 327,676 \\
MPMCT gate & 1 & 6 & 27 & 108 & 405 \\
Transposition & 1 & 6 & 28 & 120 & 496 \\
\bottomrule
  \end{tabular}
\end{table*}

In this section we are counting self-inverse functions and subclasses of them.
All results are summarized in Table~\ref{tab:counting} which also has a row for
all reversible functions as a baseline for comparison.  There are $2^n!$
reversible functions over $n$ variables due to the one-to-one correspondence
with elements in $S_{2^n}$.

Self-inverse functions over $n$ variables are characterized by their type which
is an integer partition of $2^n$.  In order to count self-inverse functions we
exploit properties from integer partitions.  Let $\mu$ be an integer partition
that contains $a_1$ ones, $a_2$ twos, and so on.  Then we define
\begin{equation}
  \label{eq:zmu}
  z_\mu = \prod_{i=1}^n i^{a_i} \prod_{i=1}^n (a_i!).
\end{equation}

\begin{lemma}[\cite{Loehr:11}]
  \label{lemma:scc}
  For a given integer partition $\mu$ of $n$, the number of permutations
  $\pi \in S_n$ for which $\type(\pi) = \mu$ is $\frac{n!}{z_\mu}$.
\end{lemma}

Based on this lemma, we can count self-inverse functions.

\begin{theorem}
  There are
  \begin{equation}
    \label{eq:count-selfinv}
    \sum_{k=0}^{2^{n-1}}(2k-1)!!\binom{2^n}{2k}
  \end{equation}
  self-inverse reversible function on $n$ variables.
\end{theorem}
\begin{IEEEproof}
  Let $N=2^n$ and $\pi \in S_N$ be an involution, i.e., $\mu = \type(\pi)$ is an
  integer partition with $k=\size(\pi)$ occurrences of $2$ and $N-2k$
  occurrences of $1$.  According to Lemma~\ref{lemma:scc} we know that there
  exist $\frac{N!}{z_\mu}$ such involutions, i.e.,
  \begin{align*}
    \frac{N!}{z_\mu} &= \frac{N!}{1^{N-2k}2^k(N-2k)!k!} =
                       \frac{N!(2k)!}{2^k(N-2k)!k!(2k)!} \\
   &= \frac{(2k)!}{2^kk!} \frac{N!}{(N-2k)!(2k)!} = (2k-1)!!\binom{N}{2k}
  \end{align*}
  The value of $k$ is bounded by $0$ and $2^{n-1}$.
\end{IEEEproof}

From~\eqref{eq:count-selfinv} we can deduce
\[ |I_n| = \sum_{k=1}^n (2^k - 1)!!\binom{2^n}{2^k}, \]
which we call palindromic in Table~\ref{tab:counting}.  The next section
determines them as the exact set of involutions that can be realized as
palindromic circuit.

We are now considering the subset of self-inverse functions that are represented
by one single-target gate.  As described above, there are $n \cdot 2^{2^{n-1}}$
single-target gates.  Single-target gates are a redundant gate representation
since $n$ gates represent the identity function, i.e., whenever the control
function is $\bot$, independent of the target line position.  Hence, the number
of functions represented by a single-target gate is
\begin{equation}
  \label{eq:count-stg}
  n \cdot 2^{2^{n-1}} - n + 1 = n(2^{2^{n-1}} - 1) + 1
\end{equation}

MPMCT gates are not redundant and there exist $n \cdot 3^{n-1}$ such gates for
$n$ variables.

Single transpositions are also a subclass of self-inverse functions and there
exist $2^{n-1}(2^n-1)$ transpositions~$\t(a b)$ over $n$ variables.  One can
choose from $2^n$ values for $a$ and from $2^n-1$ values for $b$.  Since
$\t(a b)=\t(b a)$, the product needs to be halved.

There are some subset relations worth to mention:
\begin{small}
\[ \text{reversible} \supset \text{self-inverse} \supset |I_n|
   \def\arraycolsep{1pt}
   \begin{array}{cl}
     \lower3pt\hbox{\rotatebox{25}{$\supset$}} & \text{single-target gate} \supset
                                        \text{MPMCT gate} \\[2pt]
     \raise3pt\hbox{\rotatebox{-25}{$\supset$}} & \text{transposition}
   \end{array}
\]
\end{small}

\section{Palindromic Circuits}
\label{sec:palindromic-circuits}
A reversible circuit $C=g_1g_2\dots g_k$ that consists of mixed-polarity
multiple-controlled Toffoli gates $g_i$, is called \emph{palindromic} if
$ g_i = g_{k+1-i}$ for all $i \in \{1,\dots,k\}$.  The circuit is called
\emph{even} if $k$ is even and \emph{odd} otherwise.

\begin{lemma}
  A palindromic circuit is even if and only if it realizes the identity
  function.
\end{lemma}

\begin{IEEEproof}
  Let $C=g_1g_2\dots g_{2k}$ be an even palindromic circuit.  From the
  definition of a palindromic circuit we have
  $g_1g_2\dots g_k = g_{2k}g_{2k-1}\dots g_{k+1}$.  Let $f$ be the function
  represented by these two subcircuits.  Then, $C$ represents the function
  $f\circ f^{-1} = \id$.

  Now let $C=g_1g_2\dots g_kg_{k+1}g_{k+2}\dots g_{2k+1}$ be an odd palindromic
  circuit.  Let $f$ be the function represented by $C$, $g$ be the function
  represented by $g_{k+1}$, and $\pi_f$ and $\pi_g$ their permutation
  representations.  According to Lemma~\ref{lem:same-type}, we have
  $\type(\pi_f) = \type(\pi_g)$.  Since $g$ has the functionality of a single
  gate we have $\type(\pi_g)\neq\type(\pi_{\rm id})$ and therefore $f\neq \id$.
\end{IEEEproof}

\begin{theorem}
  \label{theo:pc-type1}
  Let $f$ be a self-inverse function on $n$ variables and $\pi_f$ its
  permutation representation.  Then $\pi_f \in I_n$ if and only if $f$ can be
  realized by an odd palindromic circuit with $n$ lines.
\end{theorem}

\begin{IEEEproof}
  Direction `$\Rightarrow$': Let $C$ be an odd palindromic circuit that realizes
  the function $f$ with middle gate $g$.  Let $\pi_f$ and $\pi_g$ their
  permutation representations.  We have $\pi_g \in G_n \subseteq I_n$.
  According to Lemma~\ref{lem:same-type} we can imply that $\pi_f \in I_n$.

  Direction `$\Leftarrow$': Let $f$ be a self-inverse function with permutation
  representation $\pi_f$ such that $\pi_f \in I_n^k$.  Choose an arbitrary gate
  $g$ with permutation representation $\pi_g \in G_n^k$.  According to
  Lemma~\ref{lem:conj-class} we can always find a permutation $\sigma$ such that
  $\pi_f = \sigma \circ \pi_g \circ \sigma^{-1}$.  Obviously, $\pi_f$ can be
  represented by a palindromic circuit.
\end{IEEEproof}

\section{Alternative Constructions}
\label{sec:altern-constr}
Theorem~\ref{theo:pc-type1} works only for those self-inverse functions that are
in $I_n$.  We will now show two circuit constructions that allow to give
palindromic circuits for any self-inverse function.  The first construction
requires an additional line and the second construction requires semi-classical
quantum gates.

Both constructions are based on the same idea.  Let $f$ be a self-inverse
function with permutation representation $\pi_f \notin I_n$ such that there
exists a $k$ with $2^{k-1} < \size(\pi_f) < 2^k$.  Let $\pi_h$ be some
permutation with $\type(\pi_h)=\type(\pi_f)$ such that there exists a
permutation $\pi_g \in G_n$ with $\size(\pi_g)=2^k$ and
$\trans(\pi_h) \subset \trans(\pi_g)$.

\begin{example}
  For $n=3$ and $\pi_f = \t(0 1)\t(3 5)\t(2 7)$ we can choose
  $\pi_g = \t(0 4)\t(1 5)\t(2 6)\t(3 7)$ (i.e., $\T(\emptyset, x_3)$) and
  $\pi_h = \t(1 5)\t(2 6)\t(3 7)$ (i.e., the circuit in Fig.~\ref{fig:gates}).
\end{example}

\mn According to Lemma~\ref{lem:conj-class} we can always find a permutation
$\sigma$ such that $\pi_f = \sigma\circ \pi_h\circ \sigma^{-1}$, however, this
cannot be represented as a palindromic circuit because $\pi_h \notin I_n$.  The
permutation $\sigma \circ \pi_g \circ \sigma^{-1}$ can instead be represented as
a palindromic circuit, however, it does not represent the same function.  Let
$\pi_r = \pi_g\circ \pi_h$.  Since $\trans(\pi_h) \subset \trans(\pi_g)$ we have
$\trans(\pi_r) = \trans(\pi_g) \setminus \trans(\pi_h)$.  Note also that we have
$\pi_h = \pi_g \circ \pi_r = \pi_r \circ \pi_g$.  In order to represent the same
function we need to cancel the transpositions in $\trans(\pi_r)$ in the circuit
computation.

\begin{example}
  In the previous example we have $\pi_r = \t(0 4)$.
\end{example}

\begin{figure}[t]
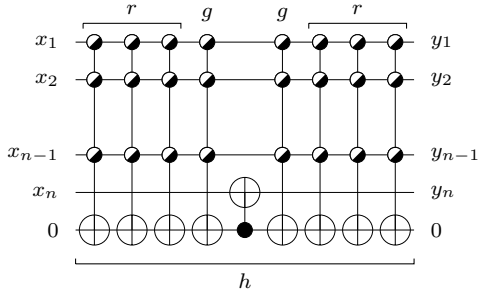

\centering
\tikzAncConstr
\caption{Construction using an additional line}
\label{fig:anc-constr}
\end{figure}
\subsection{Construction Using An Additional Line}
The construction using one additional line is depicted in
Fig.~\ref{fig:anc-constr}.  The permutation $\pi_h$ can be realized by
$\pi_r \circ \pi_g$ as described above, where $\pi_g$ is realized by a single
gate and $\pi_r$ can be realized by $\size(\pi_r)$ fully controlled Toffoli
gates.  Storing the value of that construction on a zero-intialized ancilla line
in fact computes the result of applying $\pi_h$.  The value can be used to
update the intended target line using a single controlled NOT gate.  Since all
gates in the realization of $\pi_h$ act on the same target line, they can be
arranged arbitrarily, and particularly in reverse order.  This restores the zero
value on the ancilla line.

\begin{figure}[t]
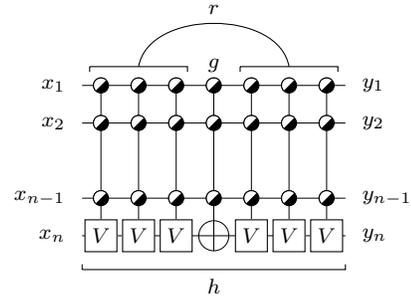

\centering
\tikzVConstr
\caption{Construction using quantum gates}
\label{fig:v-constr}
\end{figure}
\subsection{Construction Using Quantum Gates}
Instead of using an ancilla line one can also use the semi-classical $V$ gate
that performs the so-called \emph{square-root of NOT}, i.e., two consecutive
applications of a $V$ perform a NOT operation.  The circuit construction is
depicted in Fig.~\ref{fig:v-constr}.  Every assignment that triggers a
transposition in $\pi_h$ also triggers a transposition in $\pi_g$ but not in
$\pi_r$.  Hence, in that case only $\pi_g$ is performed and the target line is
updated as intended.  However, an assignment that triggers $\pi_g$ but is not in
$\pi_h$ must also trigger a transposition in $\pi_r$.  Since each of the $V$
gates are fully controlled, two of them are executed which together cancel the
update of $\pi_g$.  Due to the construction of $\pi_r$ there is no such case in
which a transposition in $\pi_r$ is triggered but not $\pi_g$.

\section{Conclusions}
\label{sec:conclusions}
In this paper we have defined palindromic circuits, a subset of the reversible
circuits, and shown the exact subclass of the self-inverse functions that can be
realized with such circuits. We have also shown how the complement (still
restricted to the self-inverse functions) to this can be constructed with either
a reversible circuit and an extra ancilla line or using quantum
gates.

To achieve the results, we investigated involutions in the symmetric group
$S_{2^n}$ that are isomorphic to self-inverse reversible functions on $n$
variables. Specifically, we define the transposition that exactly define a
reversible gate and define the rest of the reversible gates using permutation
product.

Our results provide a better understanding of the relationship between
reversible circuits and invertible functions. The understanding of this
relationship is still limited; although we only touched a subset of both areas
in this paper, we believe that this paper gives a valuable step forward.

\section*{Acknowledgement}
This work was partly funded by the \emph{European Commission} under the $7^{th}$ \emph{Framework Programme}.

\bibliographystyle{IEEEtran}
\bibliography{library}

\begin{thebibliography}{10}
\providecommand{\url}[1]{#1}
\csname url@samestyle\endcsname
\providecommand{\newblock}{\relax}
\providecommand{\bibinfo}[2]{#2}
\providecommand{\BIBentrySTDinterwordspacing}{\spaceskip=0pt\relax}
\providecommand{\BIBentryALTinterwordstretchfactor}{4}
\providecommand{\BIBentryALTinterwordspacing}{\spaceskip=\fontdimen2\font plus
\BIBentryALTinterwordstretchfactor\fontdimen3\font minus
  \fontdimen4\font\relax}
\providecommand{\BIBforeignlanguage}[2]{{%
\expandafter\ifx\csname l@#1\endcsname\relax
\typeout{** WARNING: IEEEtran.bst: No hyphenation pattern has been}%
\typeout{** loaded for the language `#1'. Using the pattern for}%
\typeout{** the default language instead.}%
\else
\language=\csname l@#1\endcsname
\fi
#2}}
\providecommand{\BIBdecl}{\relax}
\BIBdecl

\bibitem{KPF:08}
L.~A.~B. Kowada, R.~Portugal, and C.~M.~H. Figueiredo, ``Reversible
  {K}aratsuba's algorithm,'' \emph{Journal of Universal Computer Science},
  vol.~12, no.~5, pp. 499--511, 2008.

\bibitem{CDKM:05}
S.~A. Cuccaro, T.~G. Draper, S.~A. Kutin, and D.~P. Moulton, ``A new quantum
  ripple-carry addition circuit,'' \emph{arXiv:quant-ph/0410184v1}, 2005.

\bibitem{DeVos:10}
A.~De~Vos, ``Reversible computer hardware,'' \emph{Electr. Notes Theor. Comput.
  Sci.}, vol. 253, no.~6, pp. 17--22, 2010.

\bibitem{MDM:07}
D.~Maslov, G.~W. Dueck, and D.~M. Miller, ``Techniques for the synthesis of
  reversible {Toffoli} networks,'' \emph{{ACM} Trans. Design Autom. Electr.
  Syst.}, vol.~12, no.~4, 2007.

\bibitem{GWDD:09}
D.~Gro{\ss}e, R.~Wille, G.~W. Dueck, and R.~Drechsler, ``Exact multiple-control
  {Toffoli} network synthesis with {SAT} techniques,'' \emph{{IEEE} Trans. on
  {CAD} of Integrated Circuits and Systems}, vol.~28, no.~5, pp. 703--715,
  2009.

\bibitem{SWH+:12}
M.~Soeken, R.~Wille, C.~Hilken, N.~Przigoda, and R.~Drechsler, ``Synthesis of
  reversible circuits with minimal lines for large functions,'' in
  \emph{Proceedings of the 17th Asia and South Pacific Design Automation
  Conference, {ASP-DAC} 2012, Sydney, Australia, January 30 - February 2,
  2012}, 2012, pp. 85--92.

\bibitem{SPMH:03}
V.~V. Shende, A.~K. Prasad, I.~L. Markov, and J.~P. Hayes, ``Synthesis of
  reversible logic circuits,'' \emph{{IEEE} Trans. on {CAD} of Integrated
  Circuits and Systems}, vol.~22, no.~6, pp. 710--722, 2003.

\bibitem{VR:05}
A.~De~Vos and Y.~V. Rentergem, ``Reversible computing: from mathematical group
  theory to electronical circuit experiment,'' in \emph{Proceedings of the
  Second Conference on Computing Frontiers, 2005, Ischia, Italy, May 4-6,
  2005}, 2005, pp. 35--44.

\bibitem{SVJ:99}
L.~Storme, A.~De~Vos, and G.~Jacobs, ``Group theoretical aspects of reversible
  logic gates,'' \emph{J. {UCS}}, vol.~5, no.~5, pp. 307--321, 1999.

\bibitem{ASTD:14}
N.~Abdessaied, M.~Soeken, M.~K. Thomsen, and R.~Drechsler, ``Upper bounds for
  reversible circuits based on {Young} subgroups,'' \emph{Information
  Processing Letters}, vol. 114, no.~6, pp. 282 -- 286, 2014.

\bibitem{KS:11}
P.~Kerntopf and M.~Szyprowski, ``Symmetry in reversible functions and
  circuits,'' in \emph{Int'l Workshop on Logic Synthesis}, 2011, pp. 67--73.

\bibitem{AS:03}
A.~V. Aho and K.~M. Svore, ``Compiling quantum circuits using the palindrome
  transform,'' \emph{arXiv}, vol. quant-ph/0311008, 2003.

\bibitem{Loehr:11}
N.~Loehr, \emph{Bijective Combinatorics}, 1st~ed.\hskip 1em plus 0.5em minus
  0.4em\relax Chapman \& Hall/CRC, 2011.

\end{thebibliography}

\end{document}